\input harvmac


\font\cmss=cmss10 \font\cmsss=cmss10 at 7pt
\def\IZ{\relax\ifmmode\mathchoice
{\hbox{\cmss Z\kern-.4em Z}}{\hbox{\cmss Z\kern-.4em Z}}
{\lower.9pt\hbox{\cmsss Z\kern-.4em Z}}
{\lower1.2pt\hbox{\cmsss Z\kern-.4em Z}}\else{\cmss Z\kern-.4em Z}\fi}

\def\a{\alpha} \def\b{\beta} \def\g{\gamma} \def\G{\Gamma}
\def\d{\delta}  \def\ee{\varepsilon} 
\def\z{\zeta} \def\th{\theta}  
 \def\l{\lambda}  \def\m{\mu} \def\n{\nu}
\def\cs{\xi}  \def\p{\pi}  \def\r{\rho}
\def\s{\sigma}   
 \def\f{\phi}   

\def\pa{\partial}   \def\half{{1\over
2}} \def\IR{I\!\!R}
\def\oa{${\cal O}(\a ')$}\def\oaa{${\cal O}(\a '^2)$}

\Title{
\vbox{\baselineskip12pt\hbox{MIT-CTP-2625}
\hbox{hep-th/9704157}}}
{T-Duality and Two-Loop Renormalization
Flows\footnote{*}
{\baselineskip12pt This work is supported in part by
funds provided by
the U.S. Department of Energy (D.O.E.) under cooperative
research agreement \#DF-FC02-94ER40818, and by NSF Grant
PHY-92-06867. E-mail:
{\tt haagense@ctp.mit.edu, olsen@ctp.mit.edu}.} }\vskip-0.15in

\centerline{
\vbox{\hsize3in\centerline{Peter E. Haagensen$^1$
and Kasper Olsen$^{1,2}$}}}
{\it
\centerline{$^1$Center for Theoretical Physics}\vskip-.15cm
\centerline{Laboratory for Nuclear Science}\vskip-.15cm
\centerline{Massachusetts Institute of Technology}\vskip-.15cm
\centerline{77 Massachusetts Avenue}\vskip-.15cm
\centerline{Cambridge, MA 02139, USA}}\vskip-.05cm
\centerline{and}\vskip-.05cm

{\it
\smallskip
\centerline{$^2$The Niels Bohr Institute}\vskip-.15cm
\centerline{Blegdamsvej 17}\vskip-.15cm
\centerline{DK 2100 Copenhagen}\vskip-.15cm
\centerline{DENMARK
}}

\nref\giveon{For a review and further references, see
A.~Giveon, M.~Porrati and E.~Rabinovici,
{\it Phys.~Rept.} {\bf 244} (1994) 77.}
\nref\buscher{T.H.~Buscher, {\it Phys.~Lett.} {\bf 194B}
(1987) 59; {\it Phys.~Lett.} {\bf 201B} (1988) 466.}
\nref\tseytlina{A.A.~Tseytlin, {\it Mod.~Phys.~Lett.} {\bf A6} (1991)
1721.}
\nref\panvel{J.~Panvel, {\it Phys.~Lett.} {\bf 284B} (1992) 50;
J.~Balog, P.~Forg\'acs, Z.~Horv\'ath and L.~Palla, {\tt hep-th/9601091}.}
\nref\haagensen{P.E.~Haagensen, {\it Phys.~Lett.} {\bf 382B}
(1996) 356.}
\nref\balog{J.~Balog, P.~Forg\'acs, Z.~Horv\'ath and  L.~Palla,
{\it Phys.~Lett.} {\bf 388B} (1996) 121.}
\nref\tseytlinb{A.A.~Tseytlin, {\it Phys.~Lett.} {\bf 178B}
(1986) 34; {\it Nucl.~Phys.} {\bf B294} (1987) 383.}
\nref\meissner{K.A.~Meissner and G.~Veneziano, {\it Phys.~Lett.}
{\bf 267B} (1991) 33; {\it Mod.~Phys.~Lett.} {\bf A6} (1991) 3397.}
\nref\maharana{J.~Maharana and J.H.~Schwarz, {\it Nucl.~Phys.}
{\bf B390} (1993) 3.}
\nref\tseytlinc{A.A.~Tseytlin, {\it Phys.~Lett.} {\bf 194B}
(1987) 63.}
\nref\hull{C.M.~Hull and P.K.~Townsend, {\it Nucl.~Phys.} {\bf B274}
(1986) 349.}
\nref\kiritsis{E.~Kiritsis, {\it Nucl.~Phys.} {\bf B405} (1993) 109.}
\nref\burgess{P.H.~Damgaard and P.E.~Haagensen,
{\tt cond-mat/9609242}; C.P.~Burgess and \phantom{xxx} C.A.~L\"utken,
{\tt cond-mat/9611070}.}
\nref\meiss{K.A.~Meissner, {\it Phys.~Lett.} {\bf 392B} (1997) 298.}
\nref\hewson{S.F.~Hewson and N.D.~Lambert, {\tt hep-th/9703143}.}
\nref\dorn{H.~Dorn and H.-J.~Otto, {\tt hep-th/9702018}.}


\bigskip

\centerline{\bf Abstract}\medskip

\vbox{\baselineskip12pt Manifest T-duality covariance of the
one-loop renormalization group flows is shown for a
generic bosonic sigma model with an abelian
isometry, by referring a set of previously derived
consistency conditions to the tangent space of the target.
For a restricted background, T-duality transformations
are then studied at the next order, and the ensuing
consistency conditions are found to be satisfied by
the two-loop Weyl anomaly coefficients
of the model. This represents
an extremely non-trivial test of the covariance of
renormalization group flows under T-duality, and a
stronger condition than T-duality invariance of the
string background effective action.}
\vfill

\vbox{\baselineskip12pt
\noindent{\it PACS numbers: 11.10.Hi, 11.10.Kk, 11.25.-w,
11.25.Db; Keywords:
string theory, sigma models, duality, perturbation theory.}}
\Date{04/97}

\baselineskip14pt
\newsec{Introduction}\bigskip

When one thinks about symmetries in quantum field theory,
the examples that are likely to come to mind are typically
of transformations which act on the fields of a theory: for
instance, gauge or flavor symmetries, or yet charge conjugation,
parity, and time reversal. Less commonly, some theories
might also possess symmetries that act on their parameter space.
The prototype of such a symmetry occurs actually not in field
theory but in lattice spin or gauge systems, where it is known as
Kramers-Wannier duality (or a generalization thereof), and it
states that some system at low temperature is
equivalent -- or dual -- to some other system at high
temperature.

One reason such symmetries are interesting is the fact that they
act on the same (parameter) space in which also acts another important
and ubiquitous symmetry: the renormalization group (RG).
Similarly to the Kramers-Wannier dualities mentioned above, 
the RG represents
transformations in the parameter space of a theory that
leave a partition function invariant. Seen in this light, then, it
becomes natural to investigate the interplay between the RG
and other symmetries acting on the parameter space of a theory.

In the context of quantum field theory one such symmetry 
is target space duality (T-duality, for short) \giveon ,
present in $d=2$ nonlinear sigma models with a target
abelian isometry (we will furthermore restrict ourselves to bosonic
models on closed worldsheets). In its simplest incarnation, it
relates the partition function for a bosonic string compactified
on a torus of radius $R$ to the same partition function evaluated
at radius $\a '/R$. In general, however, in order to identify
a T-duality symmetry (at least to lowest nontrivial order in $\a '$,
in which case we refer to it as {\it classical} duality
symmetry), all that is required is that the target possess an abelian
isometry, so that actually a large class of targets enjoys such
a symmetry.

In spite of its being inevitably tied to the
language of string theory, it is clear 
that T-duality symmetry presents an interest quite
independently of its relevance for string theory. It is with
this interest in mind that we approach the issue of two-loop corrections
to duality transformations.

The easiest path to T-duality transformations is probably the one
given by Buscher \buscher , in which the path integral is considered
for a generic sigma model with an abelian target isometry. The isometry is
gauged, and this leads to two possible background descriptions of the
same path integral, obtained by performing the trivial parts of the path
integral in a different order. This derivation, valid at first
nontrivial order (\oa ), is classic by now, and we will not
reproduce it. The fields which do not undergo this gauging procedure
simply play a spectator role and, apart from a Jacobian factor
which appears in the course of performing these integrations
(leading to a dilaton shift in the transformations),
the procedure is simple enough that one may imagine there should not
be any significant difference at higher orders. 
There emerges the vague expectation that while one might obtain
more complicated Jacobian factors, leading to more complicated
dilaton shifts, classical duality transformations
should otherwise remain just as good a symmetry.

Unfortunately this expectation does not seem to be realized.
In a related development,
Tseytlin found in \tseytlina\ that for a particular set of backgrounds,
classical duality transformations did {\it not} keep the two-loop
string background effective action invariant, and showed
what the needed corrections to the transformations were
so that the effective action would indeed be invariant at that order.
Even in that restrictive case the necessary two-loop corrections
did not amount to any field redefinition arising from a target
reparametrization. Other authors have also noticed a similar
breakdown of classical duality symmetry \panvel .

Independently of the existence at all of a string background
effective action, but rather from a purely 2d field theoretical
point of view, it was found in \haagensen\ that the requirement of
consistency between the sigma model RG flow and classical duality
transformations imposed stringent conditions on the beta functions
of the model. For a generic background, these conditions are
satisfied by, and only by, the correct and well-known one-loop
beta functions. It is this
consistency which concerns us here, and we are led
to the main question we shall address: what do possible two-loop
corrections to duality transformations entail for the
consistency between T-duality symmetry and the RG?

In Section 2, we phrase this consistency requirement as the
commutation of two motions in the space of the theory, defined
through the action of T-duality and through the Weyl anomaly
coefficients, respectively. We then show that the consistency
conditions presented in \haagensen , which appear complicated and
unwieldy, can be stated in a
compact expression, by referring the Weyl anomaly coefficients
to the tangent space of the target. This expresses the manifest
covariance of the RG flow under the duality group $\IZ_2$. We then
examine how this may be modified at higher order.

While classical duality transformations may not be a symmetry of the
two-loop string effective action, the indication so far is that
local perturbative modifications of these transformations
do exist such that a T-duality symmetry can still be defined at higher
orders in $\a '$.
In Section 3, we reestablish contact with the string background
effective action, considering modified duality transformations at
higher order, and we analyze to what extent the consistency conditions
may or may not imply duality invariance of the background effective
action and vice-versa. If the duality transformations themselves are
modified at next order, that will induce modifications in the consistency
conditions, and {\it a priori} there is no telling what may happen
with the relation between the two-loop RG flow and the two-loop
duality transformations. Some authors \balog\
have suggested that the simple connection
between beta functions and duality transformations found at
order $\a '$ breaks down, and that in fact
T-duality at two loops does not even map sigma models into
sigma models. Given the particularly neat and compact
expression of RG covariance under one-loop duality shown in Section 2,
it would be quite disappointing to see this entire structure
simply dismantled at the next order. In Section 4 we consider
a particular class of backgrounds and show that,
quite on the contrary, the
consistency between duality and the RG remains alive and well
at two-loop order, in that the two-loop beta functions
perfectly fit the consistency conditions engendered by duality
at two loops, and in particular fixed points of one theory are
mapped to fixed points of its dual.

In Section 5 we present some conclusions and outlook, and in the
Appendix we collect the formulas summarizing the Kaluza-Klein
reduction relevant to the verification of the consistency conditions.

\newsec{Manifest Covariance at One-Loop Order}\medskip

We consider a $d\! =\! 2$ bosonic sigma model
on a generic $(D\! +\! 1)$-dimensional background
of metric $g_{\m\n}(X)$ and antisymmetric tensor $b_{\m\n}(X)$, with the
requirement that it have an abelian isometry in one target direction.
Furthermore, we choose (adapted) target coordinates $X^{\m}=(\th ,X^i)$
such that the isometry lies in the $\th$-direction. In these coordinates,
all background fields depend only on $X^i$.
The sigma model action is
\eqn\original{\eqalign{S={1\over 4\p\a '}\int d^2\!\s\, &\left[
g_{00}(X) \pa_\a \th\pa^\a\th +2g_{0i}(X)\pa_\a\th\pa^\a
X^i+g_{ij}(X)\pa_\a X^i\pa^\a X^j+ \right.\cr &\left. i\ee^{\a\b}\left(
2b_{0i}(X)\pa_\a\th\pa_\b X^i+b_{ij}(X)\pa_\a X^i \pa_\b
X^j\right)\right]\, .}}

With proper regularization and renormalization in place,
renormalized background couplings become functions
of a subtraction scale $\m$, so that they encode the RG flow of the
model through their dependence on $\m$.
Classical duality transformations are given by:
\eqn\duality{\eqalign{\tilde{g}_{00}&={1\over g_{00}}\ ,\cr
\tilde{g}_{0i}&={b_{0i}\over g_{00}}\ ,\ \ \ \tilde{b}_{0i}={g_{0i}\over
g_{00}}\ ,\cr \tilde{g}_{ij}&=g_{ij}-{g_{0i}g_{0j}-b_{0i}b_{0j}\over g_{00}}
\ ,\cr
\tilde{b}_{ij}&=b_{ij}-{g_{0i}b_{0j}-b_{0i}g_{0j}\over g_{00}}\ ,}}
mapping the background $\{ g_{\m\n},b_{\m\n}\}$ onto a dual
background $\{ \tilde{g}_{\m\n},
\tilde{b}_{\m\n}\}$. On a curved worldsheet another background coupling
is required to ensure renormalizability, that of the dilaton $\f (X)$ to
the worldsheet scalar curvature. In our approach \haagensen  , we
initially leave
the dilaton transformation under duality unspecified until consistency
conditions are enforced, at which point it becomes uniquely determined
to be
\eqn\diltransf{\tilde{\f}=\f-\half\ln g_{00}\ .}

The RG flow of the background couplings is given by their respective
beta functions:
\eqn\betas{\b_{\m\n}^g\equiv \m {d\over d\m}g_{\m\n}\
,\ \ \ \b_{\m\n}^b\equiv \m {d\over d\m}b_{\m\n}\ ,
\ \ \ \b^{\f}\equiv \m {d\over d\m}\f\ ,}
while the unintegrated (worldsheet)
stress-energy trace, which yields the conformal
anomaly of the model, is determined from the Weyl anomaly coefficients,
given by \tseytlinb :
\eqn\weyl{\eqalign{\bar{\b}_{\m\n}^g=&\b_{\m\n}^g+2\a '\nabla_\m
\pa_\n\f\cr \bar{\b}_{\m\n}^b=&\b_{\m\n}^b+\a 'H_{\m\n}^{\ \
\l}\pa_\l\f\cr \bar{\b}^\f=&\b^\f+\a '(\pa_\m\f)^2 \ .}}

In previous work \haagensen , the consistency conditions
to be presented below
were found to hold at lowest order both for the beta functions and the Weyl
anomaly coefficients. However, while they are satisfied up to a target
reparametrization by the beta functions (which is reasonable to expect),
they are on the other hand {\it identically} satisfied by the Weyl
anomaly coefficients (a deep reason for which, from the 2d field theory
point of view, we have not found). Thus,
although strictly speaking the Weyl anomaly coefficients do not represent
an RG motion in the parameter space, in order to be concise, we will
mainly be referring to these coefficients in what follows, making the
distinction from the beta functions when necessary.

We now define an operation $R$, akin to the RG motion $\m {d\over d\m}$:
\eqn\R{R\pmatrix{g_{\m\n}\cr b_{\m\n}\cr \f}=
\pmatrix{\bar{\b}^g_{\m\n}[g,b,\f ]\cr \bar{\b}^b_{\m\n}[g,b,\f ]
\cr \bar{\b}^\f [g,\f ]}\ ,}
so that on a generic functional $F[g,b,\f]$,
\eqn\RF{RF[g,b,\f]={\d F\over\d g_{\m\n}}\cdot\bar{\b}_{\m\n}^g
+{\d F\over\d b_{\m\n}}\cdot\bar{\b}_{\m\n}^b
+{\d F\over\d \f}\cdot\bar{\b}^\f\ ,}
and a duality operation $T$:
\eqn\T{T\pmatrix{g_{\m\n}\cr b_{\m\n}\cr \f}=
\pmatrix{\tilde{g}_{\m\n}[g,b]\cr \tilde{b}_{\m\n}[g,b]
\cr \tilde{\f}[g,\f ]}\ ,}
affecting the duality transformations \duality , \diltransf\
and, more generally,
\eqn\TF{TF[g,b,\f]=F[\tilde{g},\tilde{b},\tilde{\f}]\ .}
At any given order in $\a '$, $R$ is defined by the corresponding
Weyl anomaly coefficients, while {\it ab initio} we only know $T$ at
lowest order, where it is given by \duality\ and \diltransf .
Whether and how $T$ is modified at higher orders is one of the crucial
questions at hand.

Then, regardless of the order at which $R$ is defined,
{\it if} the action of $T$ on $g_{\m\n}$ and $b_{\m\n}$
is defined by \duality , the
requirement that these two motions in the space of the theory
commute: \eqn\TR{[T,R]=0\ ,} can be seen through \RF\ and \TF\ to be
tantamount to the following consistency conditions \haagensen :
\eqn\consistency{\eqalign{ \bar{\b}^{\tilde{g}}_{00}&=-{1\over g_{00}^2}
\bar{\b}^g_{00}\ ,\cr \bar{\b}^{\tilde{g}}_{0i}&=-{1\over g_{00}^2}\left(
b_{0i}\bar{\b}^g_{00}-\bar{\b}^b_{0i}g_{00} \right)\ , \cr
\bar{\b}^{\tilde{b}}_{0i}&=-{1\over g_{00}^2}\left(
g_{0i}\bar{\b}^g_{00}-\bar{\b}^g_{0i}g_{00} \right)\ ,\cr
\bar{\b}^{\tilde{g}}_{ij}&=\bar{\b}^g_{ij}-{1\over g_{00}}\left(
\bar{\b}^g_{0i}g_{0j}+
\bar{\b}^g_{0j}g_{0i}-\bar{\b}^b_{0i}b_{0j}-\bar{\b}^b_{0j}b_{0i}\right)
+ {1\over
g_{00}^2}\left( g_{0i}g_{0j}-b_{0i}b_{0j}\right) \bar{\b}^g_{00}\ ,\cr
\bar{\b}^{\tilde{b}}_{ij}&=\bar{\b}^b_{ij}-{1\over g_{00}}\left(
\bar{\b}^g_{0i}b_{0j}+
\bar{\b}^b_{0j}g_{0i}-\bar{\b}^g_{0j}b_{0i}-\bar{\b}^b_{0i}g_{0j}\right)
+ {1\over
g_{00}^2}\left( g_{0i}b_{0j}-b_{0i}g_{0j}\right) \bar{\b}^g_{00}\ .}}

As shown in \haagensen , enforcing these conditions at ${\cal O}(\a ')$
uniquely determines the beta functions at that order
to be (up to a global factor):
\eqn\betaone{\eqalign{\b_{\m\n}^g=&\a '\left( R_{\m\n} -{1\over4}
H_{\m\l\r} H_{\n}^{\ \l\r}\right)\ ,
\cr \b_{\m\n}^b=&-{\a '\over2}\nabla_\l
H^\l_{\ \m\n}\ ,}}
where $H_{\m\n\l}=\pa_\m b_{\n\l}+{\rm cyclic\
permutations}$, and the dilaton transformation, or ``shift'', to be
given by \diltransf . Furthermore, one can now apply the same
condition $[T,R]=0$ to the dilaton transformation, obtaining
yet another consistency condition:
\eqn\dilconsist{\bar{\b}^{\tilde{\f}} =\bar{\b}^\f
-\half {1\over g_{00}}\bar{\b}^g_{00}\ .}
This is satisfied for
\eqn\dilbetaone{\b^\f=C-{\a '\over2}\nabla^2\f\ ,}
with $C$ an arbitrary constant, so that all beta functions are
determined at one-loop order up to a global factor and the value of $C$ .

For a model with $n$ abelian isometries, one expects the duality
symmetry to be $O(n,n,\IZ )$. (In the context of background effective
actions, where the sigma model abelian isometry leads to spontaneous
compactification, this was nicely shown in \meissner\ and in \maharana .
The connection with the sigma model context is also explored in these
and related works.)
In our case, this is just $\IZ_2$ ($O(1,1,\IR )$ is the group
of hyperbolic rotations on the plane; restricting it to matrix
representatives with integer entries eliminates all group elements but
$\pm \rlap1\kern.15em 1$).
It is difficult to imagine that covariance under
such a simple symmetry as $\IZ_2$ cannot be expressed in terms simpler
than \consistency . We now show that this is in fact possible if
the tensors in  \consistency\ are referred to the tangent frame of the
target.{\footnote{$^{\dag}$}{That the consistency conditions would simplify
when referred to the tangent space was first noticed by P.~Letourneau
(private communication).}

As in \haagensen , we decompose the generic metric $g_{\m\n}$
as follows:
\eqn\metric{g_{\m\n}=\pmatrix{a & av_i\cr av_i &
\bar{g}_{ij} +av_iv_j}\ ,}
so that $g_{00}\!=\!a,\ g_{0i}\!=\!av_i,
\ g_{ij}\!=\!\bar{g}_{ij}+av_iv_j$.
The components of the antisymmetric tensor are written
as $b_{0i}\equiv w_i$ and $b_{ij}$. From \duality\ we find that
in terms of this decomposition,
the dual metric and antisymmetric tensor are given by the
substitutions $a\rightarrow 1/a, v_i\leftrightarrow w_i$, and
$\tilde{b}_{ij}\!=\!b_{ij}+w_iv_j-w_jv_i$. The vielbeins
corresponding to \metric\ can always be taken in the block
triangular form:
\eqn\bein{e_{\m}^{\ a}=\pmatrix{e_{0}^{\ \hat{0}} &e_{0}^{\ \a} \cr
e_{i}^{\ \hat{0}} &e_{i}^{\ \a} }
=\pmatrix{\sqrt{a} & 0\cr\sqrt{a} v_i &
\bar{e}_{i}^{\ \a}\ }\ ,}
where tangent space indices are decomposed as $a=\hat{0},\a$;
$\a=1,2,\ldots ,D$  (corresponding to the decomposition $\m=0,i$;
$i=1,2,\ldots ,D$), and $\bar{e}_{i}^{\ \a}\bar{e}_{j}^{\ \b}\d_{\a\b}
=\bar{g}_{ij}$.
For ease of reference, we also present here the inverse vielbein:
\eqn\beininverse{e_{a}^{\ \m}=\pmatrix{e_{\hat{0}}^{\ 0}
&e_{\hat{0}}^{\ i} \cr
e_{\a}^{\ 0} &e_{\a}^{\ i} }
=\pmatrix{1/\sqrt{a} & 0\cr -v_\a &
\bar{e}_{\a}^{\ i}\ }\ ,}
with $v_\a\equiv\bar{e}_{\a}^{\ i}v_i$.

The tangent space Weyl anomaly coefficients are defined through:
\eqn\Beta{\bar{\b}^{g}_{a b}=e_{a}^{\ \m}e_{b}^{\ \n}
\bar{\b}^{g}_{\m\n}\ \ ,\ \ \bar{\b}^{b}_{a b}=e_{a}^{\ \m}
e_{b}^{\ \n} \bar{\b}^{b}_{\m\n}\ ,}
while an analogous definition holds in the dual background:
\eqn\Beta{\bar{\b}^{\tilde{g}}_{a b}=
\tilde{e}_{a}^{\ \m}\tilde{e}_{b}^{\ \n}
\bar{\b}^{\tilde{g}}_{\m\n}\ \ ,\ \
\bar{\b}^{\tilde{b}}_{a b}=\tilde{e}_{a}^{\ \m}
\tilde{e}_{b}^{\ \n} \bar{\b}^{\tilde{b}}_{\m\n}\ .}

Using \consistency , it is now straightforward to work out the
appropriate consistency relations for the tangent space anomaly
coefficients. For
illustration purposes, we show here the $\hat{0}\hat{0}$ and
$\hat{0}\a$ components:
\eqn\betann{
\bar{\b}^{\tilde{g}}_{\hat{0}\hat{0}}=
\tilde{e}_{\hat{0}}^{\ 0}\ \tilde{e}_{\hat{0}}^{\ 0}\
\bar{\b}^{\tilde{g}}_{00}=a\ \bar{\b}^{\tilde{g}}_{00}=
-{1\over a}\ \bar{\b}^{g}_{00}=-e_{\hat{0}}^{\ 0}\ e_{\hat{0}}^{\ 0}\
\bar{\b}^{g}_{00}=-\bar{\b}^{g}_{\hat{0}\hat{0}}\ ,}

\eqn\betana{\eqalign{
\bar{\b}^{\tilde{g}}_{\hat{0}\a}&=\tilde{e}_{\hat{0}}^{\ 0}\
\tilde{e}_{\a}^{\ 0}\ \bar{\b}^{\tilde{g}}_{00}+
\tilde{e}_{\hat{0}}^{\ 0}\ \tilde{e}_{\a}^{\ i}\
\bar{\b}^{\tilde{g}}_{0i}\cr
&=-\sqrt{a}\ w_j\bar{e}_{\a}^{\ j}\bar{\b}^{\tilde{g}}_{00}
+\sqrt{a}\ \bar{e}_{\a}^{\ i}\bar{\b}^{\tilde{g}}_{0i}\cr
&= {1\over a^2}\sqrt{a}\ w_j\bar{e}_{\a}^{\ j}\bar{\b}^{g}_{00}
-{1\over a^2}\sqrt{a}\ \bar{e}_{\a}^{\ i}\left( w_i\bar{\b}^{g}_{00}
-a\bar{\b}^{b}_{0i}\right)
\cr
&= e_{\hat{0}}^{\ 0}\ e_{\a}^{\ i}\ \bar{\b}^b_{0i}=
\bar{\b}^{b}_{\hat{0}\a}\ .}}

The entire set of consistency conditions reads:
\eqn\Bconsistency{\eqalign{\bar{\b}^{\tilde{g}}_{\hat{0}\hat{0}}
=&-\bar{\b}^g_{\hat{0}\hat{0}}\ ,\cr
\bar{\b}^{\tilde{g}}_{\hat{0} \a}= \bar{\b}^{b}_{\hat{0} \a}\ \ ,&\ \
\bar{\b}^{\tilde{b}}_{\hat{0} \a}= \bar{\b}^{g}_{\hat{0} \a}\ ,\cr
\bar{\b}^{\tilde{g}}_{\a\b}= \bar{\b}^{g}_{\a\b}\ \ ,&\ \
\bar{\b}^{\tilde{b}}_{\a\b}= \bar{\b}^{b}_{\a\b}\ ,}}
or, in a slightly more compact form:
\eqn\conscomp{\eqalign{
\left( \bar{\b}^{\tilde{g}}\pm \bar{\b}^{\tilde{b}}
\right)_{\hat{0}\hat{0}} & =-
\left( \bar{\b}^g\pm \bar{\b}^b\right)_{\hat{0}\hat{0}}\ ,\cr
\left( \bar{\b}^{\tilde{g}}\pm \bar{\b}^{\tilde{b}}
\right)_{\hat{0}\a} & =\pm
\left( \bar{\b}^g\pm \bar{\b}^b\right)_{\hat{0}\a}\ ,\cr
\left( \bar{\b}^{\tilde{g}}\pm \bar{\b}^{\tilde{b}}
\right)_{\a\b} & = +
\left( \bar{\b}^g\pm \bar{\b}^b\right)_{\a\b}
\ .}}
In either form, the $\IZ_2$ duality covariance is now manifestly
seen. Furthermore, the job of actually verifying the consistency
relations, which was rather lengthy in original form \haagensen ,
is also found to simplify. This happens due to the
fact that the Kaluza-Klein reduction of the relevant geometric tensors
``regroups'' into much simpler structures. This can be seen
from the expressions in the Appendix for $R_{\hat{0}\hat{0}},
R_{\hat{0}\a}$ and $R_{\a \b}$ as opposed to $R_{00},R_{0i}$
and $R_{ij}$. The same can be also seen, in fact, for the expressions
for the Riemann tensor, so that at higher orders this simplification
should continue to occur.

The analogous conditions for
the beta functions differ from the above by a target diffeomorphism
\haagensen , which however can also be easily stated in the tangent
frame:
\eqn\name{\eqalign{\b^{\tilde{g}}_{\hat{0}\hat{0}}=-\b^g_{\hat{0}\hat{0}}
&+\a'\nabla_{(\hat{0}}\cs_{\hat{0})}\ ,\cr
\b^{\tilde{g}}_{\hat{0} \a}= \b^{b}_{\hat{0} \a}
-\a'H_{\hat{0}\a}^{\ \ \g}\cs_{\g}\ \ ,&\ \
\b^{\tilde{b}}_{\hat{0} \a}= \b^{g}_{\hat{0} \a}
-\a'\nabla_{(\hat{0}}\cs_{\a)}\ ,\cr
\b^{\tilde{g}}_{\a\b}= \b^{g}_{\a\b}
-\a'\nabla_{(\a}\cs_{\b)}\ \ ,&\ \
\b^{\tilde{b}}_{\a\b}= \b^{b}_{\a\b}
-\a'H_{\a\b}^{\ \ \g}\cs_{\g}\ , }}
where $\cs_{a}=-\half e_{a}^{\ \m}\pa_{\m}\ln g_{00}$, and $(ab)=ab+ba$.

Given the extremely
simple form of the above consistency conditions, one might na\"{\i}vely
hope that this structure would not change at higher orders.
However, if one views the consistency conditions as stated through
the commutator $[T,R]=0$, it then becomes apparent that it would be
very unlikely that T-duality transformations would not change: in
going one order higher, the operator $R$ is modified by the next
order beta functions, succinctly, $R\equiv \a 'R_1
\to \a 'R_1+\a '^2R_2$, in obvious notation. The condition
to be demanded for consistency, at any order, should be that the
$R$ and $T$ motions commute, so that one should in fact expect
$T\equiv T_1\to T_1+\a 'T_2$ as well. This, however, should not detract
from the highly nontrivial statement that a $T$ operation can be defined
at all such that $[T,R]=0$ at higher orders.
In Section 4, we will investigate this for a restricted class of
backgrounds.

\newsec{Consistency Conditions and Duality Invariance}\medskip

We consider in this Section to what extent
duality invariance and the consistency conditions may imply
each other, and general requisites to be expected of
a modified T-duality transformation at two-loop order if it is to leave
the background effective action invariant. For this, it is useful to
note that the integrand of the background effective action,
\eqn\ea{L_{\rm eff}=\sqrt{g}e^{-2\f}\left(\bar{\b}^\f-{1\over4}g^{\m\n}
\bar{\b}^g_{\m\n}\right)\ ,}
can be obtained by the $R$ operation acting on the ``measure'' factor
$V\equiv\sqrt{g}e^{-2\f}$ \tseytlinc :
\eqn\rvol{-\half R\left(\sqrt{g}e^{-2\f}\right)=
\sqrt{g}e^{-2\f}\left(\bar{\b}^\f-{1\over4}g^{\m\n}
\bar{\b}^g_{\m\n}\right)\ .}
This is fairly simple to verify by using the fact that $\sqrt{g}
\equiv\sqrt{\det g}=\sqrt{\det \bar{g}}\sqrt{g_{00}}$, and is
valid at higher orders and for generic backgrounds.

From this, it becomes clear that a way to achieve
T-invariance of the effective action at some higher order
is to require, beyond
the commutation of $T$ and $R$, that $V$ be invariant under $T$, for
then:
\eqn\eainvone{TR\left(\sqrt{g}e^{-2\f}\right)=
T\left[\sqrt{g}e^{-2\f}\left(\bar{\b}^\f-{1\over4}g^{\m\n}
\bar{\b}^g_{\m\n}\right)\right]=
\sqrt{\tilde{g}}e^{-2\tilde{\f}}\left(\bar{\b}^{\tilde{\f}}-
{1\over4}\tilde{g}^{\m\n}\bar{\b}^{\tilde{g}}_{\m\n}\right)\ ,}
while
\eqn\eainvtwo{R\ T\left(\sqrt{g}e^{-2\f}\right)=
R\left(\sqrt{g}e^{-2\f}\right)=
\sqrt{g}e^{-2\f}\left(\bar{\b}^\f-{1\over4}g^{\m\n}
\bar{\b}^g_{\m\n}\right)\ .}
Thus, $[T,R]=0$, together with the T-invariance of $V$ implies
that the background effective action is also T-invariant. Similar
reasoning also shows that, conversely, T-invariance of both $V$ and
the effective action implies that the commutator $[T,R]$ also vanishes
when acting on $V$. This, naturally, is a weaker statement than
$[T,R]=0$ as an operator identity. A simple example shows this clearly:
if we take $T$ to be the usual one-loop duality transformations, and
$R$ to be the map into the \oa \ Weyl anomaly coefficients, except for
$\bar{\b}^b_{\m\n}$, which we take to be wrong, say, twice the
correct value, then $TV=V$, $TL_{\rm eff}=L_{\rm eff}$, and $[T,R]V=0$.
Yet, $[T,R]\neq 0$, as the consistency conditions are not satisfied
(and it is not the case that the consistency conditions for
$\bar{\b}^b_{\m\n}$ ``decouple'' from the invariance of the
effective action, because they contain $\bar{\b}^g_{\m\n}$'s, which
are present in the effective action).

The above does not preclude the possibility that T-invariance of
the effective action may be achieved without the invariance of $V$;
however, a more detailed examination of the specific terms involved
at \oaa\ shows that this possibility is considerably more
complicated, so that we will choose to discard it while we can (and
at \oaa\ we can). The corrections to $T$ that preserve the invariance
of $V$ are rather easily found: if we assume them to be
\eqn\tcorr{\eqalign{
\ln\tilde{g}_{00}=&-\ln g_{00}+2\a 'Q_0\ ,\cr
\tilde{\f}=& \f -\half\ln g_{00}+\a 'Q_\f}}
as well as other (for now unimportant) corrections on the
remaining background fields, then
\eqn\voltrsf{\eqalign{
\sqrt{g}e^{-2\f}=&\sqrt{\det\bar{g}}\sqrt{g_{00}}e^{-2\f}
\buildrel T\over{\hbox to 1cm{\rightarrowfill}}
\sqrt{\det\bar{g}}\sqrt{\tilde{g}_{00}}
e^{-2\tilde{\f}}\cr
=&\sqrt{\det\bar{g}}\sqrt{g_{00}}e^{-2\f +\a '(Q_0-2Q_\f)}\ .}}
Thus, for any $Q_0$ and $Q_\f$ satisfying $Q_0=2Q_\f$, $V$ will
be T-invariant. Of course, one must now verify whether
any such corrections exist at all so that also $[T,R]=0$ is satisfied.
This will be done in the next section, for a particular class
of backgrounds.

To summarize, the
following statements hold:
\item{{\it i})} $[T,R]=0$ does {\it not} imply $TL_{\rm eff}=L_{\rm eff}$;
\item{{\it ii})} $TL_{\rm eff}=L_{\rm eff}$ does {\it not} imply $[T,R]=0$;
\item{{\it iii})} $[T,R]=0$ {\it and} $TV=V$ does imply
$TL_{\rm eff}=L_{\rm eff}$;
\item{{\it iv})} $TL_{\rm eff}=L_{\rm eff}$ {\it and} $TV=V$
implies $[T,R]V=0$, but does {\it not} imply $[T,R]=0$ in general.

The requirement motivated by string theory
is that the background effective action should be T-invariant. In
light of the above considerations, however, we would instead
elevate to a basic principle the requirement of consistency
between duality and the RG flow in the sigma model, $[T,R]=0$.
Then, in order to furthermore achieve duality invariance of the background
effective action in the simplest way, one should also impose the
T-invariance of $V\equiv\sqrt{g}e^{-2\f}$.

\newsec{Covariance at Two-Loop Order}\medskip

A simple glance at the two-loop sigma model beta functions
is sufficient to convince one that
consistency conditions at \oaa\ for a generic background are
extremely complicated. We choose instead to work on a more
restricted background, where we will nevertheless be able to
illustrate in a highly nontrivial way how consistency conditions
are satisfied. We take a background previously considered by
Tseytlin \tseytlina :
\eqn\metricres{g_{\m\n}=\pmatrix{a & 0\cr 0 &\bar{g}_{ij}}\ ,}
and $b_{\m\n}=0$. Two sets of corrections to duality transformations
were found in \tseytlina\ such that the effective action
remains invariant at two-loop order. It turns out that
only one of these will furthermore satisfy the consistency conditions.
These corrected transformations are:
\eqn\dualitymodified{\eqalign{
\ln\tilde{a}=&-\ln a+{\a'\over2}a_{i}a^{i}\ ,
\ \ \ \tilde{g}_{ij}=g_{ij}=\bar{g}_{ij}\cr
\tilde{\f}=&\f -{1\over2}\ln a +{\a'\over8}a_{i}a^{i}\ ,}}
where $a_{i}\equiv\pa_i\ln a$, and indices $i,j,\ldots$ are raised
with the inverse metric $\bar{g}^{ij}=(\bar{g}_{ij})^{-1}$
(cf. the Appendix for further details).
In particular, these modified transformations satisfy the condition
$Q_0=2Q_\f$ stated previously, so that $V$ remains T-invariant.
Consistency conditions follow from applying $R$ to \dualitymodified\
(and using $[T,R]=0$ on the l.h.s.), leading to:
\eqn\consistencymodified{\eqalign{
{1\over \tilde{a}}\ \tilde{\bar{\b}}_{00} =&
-{1\over a}\ \bar{\b}_{00}+\a'\left[
a^{i}\pa_{i}\left( {1\over a}\ \bar{\b}_{00}\right)
-\half a^{i}a^{j}\bar{\b}_{ij}\right]\ ,\cr
\tilde{\bar{\b}}_{ij}= & \ \bar{\b}_{ij}\ ,\cr
\tilde{\bar{\b}}^{\f}=& \ \bar{\b}^\f-{1\over 2a}\ \bar{\b}_{00}
+{\a'\over4}\left[ a^{i}\pa_{i}\left( {1\over a}\ \bar{\b}_{00}\right)
-\half a^{i}a^{j}\bar{\b}_{ij}\right]\ .}}
The anomaly coefficients appearing inside the square brackets
should only be taken to \oa , as the equations are valid
to \oaa .
It now becomes manifest that while the above conditions
certainly imply the invariance of $L_{\rm eff}$, the converse is not true.
However, we can use the established invariance of $L_{\rm eff}$ under
\dualitymodified\ to our advantage, insofar as it implies
that once we have shown the first two consistency conditions to hold,
the third
one has no other option but to be satisfied. The rest of this section
is devoted to proving the first two consistency conditions in
\consistencymodified .

The \oaa\ Weyl anomaly coefficients of the model are \tseytlinb :
\eqn\betafns{\eqalign{
\bar{\b}_{\m\n}= &\  \a '(R_{\m\n}+2\nabla_\m\pa_\n\f )+
{\a '^2\over2}R_\m^{\ \l\r\s}R_{\n\l\r\s}\ ,\cr
\bar{\b}^\f= & \ {D-25\over6}-{\a '\over2}(\nabla^2\f-2\pa_\m\f\pa^\m\f )+
{\a '^2\over16}R^{\m\l\r\s}R_{\m\l\r\s}\ .}}
Using the Kaluza-Klein reduction formulas found in the Appendix, the
metric anomaly coefficients translate into:
\eqn\betakk{\eqalign{
{1\over a}\ \bar{\b}_{00}=&\  {\a'\over2}\left( -q_{i}^{\ i}
+2a^i\pa_i\f\right)+{\a'^2\over4}q_{ij}q^{ij}\ , \cr
\bar{\b}_{ij}=&\ \a'\left(\bar{R}_{ij}-\half q_{ij}
+2\bar{\nabla}_{i}\pa_{j}\f\right)
+{\a'^2\over2}\left( \bar{R}_{ikmn}\bar{R}_{j}^{\ kmn}
+\half q_{ik}q_{j}^{\ k}\right)\ ,}}
where $q_{ij}=\bar{\nabla}_ia_j+\half a_ia_j$, and $\bar{\nabla}_i,
\bar{R}_{ij}, \bar{R}_{ijmn},$ etc. refer to tensors calculated
in the reduced metric $\bar{g}_{ij}$.
Incidentally, although the expressions in the Appendix show that
for a generic background most formulas are considerably simplified
by referring them to the tangent space, in this restricted case
no great simplification is achieved with that. We therefore choose
to keep the usual indices for clarity of presentation.

Corrected duality transformations take $a\to\tilde{a}$ and
$\f\to\tilde{\f}$ as in \dualitymodified\ and, consequently,
\eqn\aq{\eqalign{
a_i\longrightarrow & \ \tilde{a_i}=-a_i+\a 'a^j\bar{\nabla}_ia_j\ ,\cr
q_{ij}\longrightarrow & \ \tilde{q}_{ij}=-q_{ij}+a_ia_j+
{\a '\over2}[\bar{\nabla}_i\pa_j-\half a_{(i}\pa_{j)}](a^ka_k)\ ,
}}
where $(ij)=ij+ji$, and we only need consider terms to \oa\ in the
duality transformations. The dual metric anomaly coefficients
are:
\eqn\dualbetakk{\eqalign{
{1\over \tilde{a}}\ \tilde{\bar{\b}}_{00}=&
\ {\a'\over2}\left( -\tilde{q}_{i}^{\ i}
+2a^i\pa_i\tilde{\f}\right)+{\a'^2\over4}\tilde{q}_{ij}
\tilde{q}^{ij}\ , \cr
\tilde{\bar{\b}}_{ij}=&\ \a'\left(\bar{R}_{ij}-\half \tilde{q}_{ij}
+2\bar{\nabla}_{i}\pa_{j}\tilde{\f}\right)
+{\a'^2\over2}\left( \bar{R}_{ikmn}\bar{R}_{j}^{\ kmn}
+\half \tilde{q}_{ik}\tilde{q}_{j}^{\ k}\right)\ .}}
Verification of the consistency conditions in
\consistencymodified\ now requires painstaking diligence, but not
too much creativity. One substitutes \aq\ into
\dualbetakk , and that and \betakk\ into \consistencymodified .
Although the entire procedure is rather long,
the only nontrivial step involves the use of the
geometrical identity
$\left[ \bar{\nabla}^{2},\bar{\nabla}_{i}\right] S
=\bar{R}_{ij}\bar{\nabla}^{j}S$, for $S$ a scalar.

The result we finally arrive at is that the consistency
conditions, \consistencymodified , are exactly satisfied, showing
that the motions $T$ and $R$ commute in the space of the
model, as we set out to demonstrate.
In \tseytlina\ another set of two-loop modified duality
transformations were found which, despite not leaving the measure
factor $V$ invariant, do leave the background effective action
invariant. This second set of duality transformations,
call it $T'$, is obtained
from \dualitymodified\ by a target diffeomorphism,
designed such as to preserve at two-loop order the one-loop relation
$\ln\tilde{a}=-\ln a$. Interestingly, we find that for $T'$, the
consistency conditions are {\it not} satisfied: $[T',R]\neq 0$.
We do not fully understand at present why the
consistency conditions are not satisfied for this second
set of duality transformations, and this curious fact may well be
worth investigating further. Nevertheless, as it has no bearing
on the result we have shown here, we will refrain from attempting
to interpret it.

Essentially all we have done until now concerns the $R$ operation
as defined through the Weyl anomaly
coefficients. Ultimately, however, the motivation underlying our
investigation rests in the requirement of consistency of duality
symmetry with true RG motions in the space of bosonic sigma models
on flat worldsheets. These RG motions are generated by beta functions,
and do not exactly coincide with the $R$ operation considered
previously, although they are of course intimately related.
We now present the consequences of what we have found
above to the beta functions of the model and its dual.

The consistency conditions \consistencymodified\ are translated into
a set of consistency conditions for the beta functions by using the fact
that beta functions differ from the Weyl anomaly coefficients
through \weyl . We first consider the $ij$ component in
\consistencymodified\ at $\tilde{\f}=0$:
\eqn\ij{\tilde{\b}_{ij}=\b_{ij}+2\a'\bar{\nabla}_{i}\pa_{j}
\left( \half\ln a -{\a'\over8}a_ia^i\right)\equiv
\b_{ij}-2\a'\bar{\nabla}_{i}\cs_{j}\ , }
with $\cs_\m=-1/2\ \pa_\m(\ln a -\a'/4\ a_ia^i)$.
Taking instead $\f\!=\!0$ in \consistencymodified , we find the same
equation to \oaa , with tilde and untilde quantities interchanged.
This is also equivalent to the more symmetric form:
\eqn\ijsymm{\tilde{\b}_{ij}-\a'\bar{\nabla}_i\tilde{\cs}_j
=\b_{ij}-\a'\bar{\nabla}_i\cs_j\ .}
These expressions represent the same consistency conditions found
in \haagensen\ for the beta functions, restricted to our particular
background, but now valid to \oaa . Something slightly different
will happen with the $00$ component, however. If we again take
$\tilde{\f}=0$, now in the first equation in \consistencymodified ,
we find:
\eqn\zerozero{
{1\over \tilde{a}}\ \tilde{\b}_{00} =
-{1\over a} \left[ \b_{00}+2\a'\nabla_0\pa_0
\left( \half\ln a -{\a'\over8}a_ia^i\right)
\right]+\a'^2\left[
a^{i}\pa_{i}\left( {\bar{\b}_{00}^{(1)}\over a}\right)
-\half a^{i}a^{j}\bar{\b}_{ij}^{(1)}\right] \ ,   }
where the superscript $(1)$ denotes the one-loop quantities
(at $\tilde{\f}=0$)
\eqn\olq{\eqalign{
\bar{\b}_{00}^{(1)} &=  \bar{\b}_{00}^{\rm (1-loop)}|_{\tilde{\f}=0}
= R_{00}+\nabla_0\pa_0\ln a\cr
&= -{a\over2}\left[ \bar{\nabla}_ia^i - \half a_ia^i\right]\ ,\cr
\bar{\b}_{ij}^{(1)} &=  \bar{\b}_{ij}^{\rm (1-loop)}|_{\tilde{\f}=0} =
{R}_{ij}+\nabla_i\pa_j\ln a\ \cr
&=\bar{R}_{ij}+\half\bar{\nabla}_ia_j-{1\over4}a_ia_j\ .}}
Using formulas from the Appendix, it is possible to see that this is
equivalent to:
\eqn\zerozerob{
{1\over \tilde{a}}\ \tilde{\b}_{00} =
-{1\over a} \left( \b_{00}-2\a'\nabla_0\cs_0-2\a'^2\nabla_0\z_0
\right)\ ,}
where
\eqn\zet{\eqalign{
\z_i=& \ \pa_i(\bar{\b}_{00}^{(1)}/a)-\half\bar{\b}_{ij}^{(1)}a^j\cr
=& \ \half\left( -\bar{\nabla}_j\bar{\nabla}_i+\half q_{ij}\right) a^j\ .}}
Like for the $ij$ components, this is equivalent to the same
equation with tilde and untilde quantities interchanged, and
it is also equivalent to the symmetric form:
\eqn\zerozerosymm{
{1\over\tilde{a}}\left[ \tilde{\b}_{00}-\a'\tilde{\nabla}_{0}
\tilde{\cs}_{0}
-\a'^2\tilde{\nabla}_{0}\tilde{\z}_{0}\right]
=-{1\over a}\left[ \b_{00}-\a'\nabla_{0}\cs_{0}
-\a'^2\nabla_{0}\z_{0}\right]\ .}
The possibility to write the equations above in these alternative forms
is of course a consequence of the fact that, even with the modifications
considered in this section, the duality transformations still correspond
to a $\IZ_2$ symmetry.

Thus, while separately both the $ij$ and $00$ components satisfy
one-loop consistency conditions up to a target diffeomorphism,
like the one-loop beta functions do, these are different diffeomorphisms
for the $ij$ and the $00$ components. This means, in particular, that
the statement that scale invariant backgrounds are mapped to scale
invariant backgrounds is still not entirely transparent. Nevertheless,
this statement is still true, since setting the two-loop beta functions
to zero, say, in the original background, leads to
\eqn\betatozero{
\beta_{\m\n}=\a'\beta_{\m\n}^{\rm (1-loop)}+\a'^2
\beta_{\m\n}^{\rm (2-loop)}=0
\Longrightarrow \beta_{\m\n}^{\rm (1-loop)}=-\a'
\beta_{\m\n}^{\rm (2-loop)}\ .}
When substituted in \zerozero , this shows that all components
of the beta functions now satisfy the one-loop consistency conditions
with the {\it same} diffeomorphism, up to terms of ${\cal O}(\a'^3)$.
This finally
implies that, to the order considered, scale invariant backgrounds
are mapped to scale invariant backgrounds \haagensen,\hull .

Thus, apart from the fact that formulas become more complicated at
\oaa , we see that nothing has gone awry, at least for the restricted
class of backgrounds presented here. Not only can a duality transformation
be defined such as to maintain the invariance of the background
effective action (as shown in \tseytlina ), but it can at the
same time be defined such as to preserve the consistency between RG flows
and duality transformations in the space of the theory. Contrary to the 
claim in \balog , we believe the dual
of a sigma model continues to be a sigma model. In particular, when
one model reaches a fixed point its dual also will.

\newsec{Conclusions}

In this paper we have studied the consistency between renormalization
group flows and T-duality symmetry in $d\! =\! 2$ bosonic
sigma models. This consistency is expressed as the requirement
that T-duality and RG flows commute, when considered as motions
in the parameter space of the theory. Such a requirement
was known to be satisfied at \oa \haagensen , where it had previously
been expressed as a complicated set of relations amongst the
Weyl anomaly coefficients of the theory. By referring these
anomaly coefficients to the target tangent space, we have been
able to considerably simplify both the expression and the verification
of consistency at \oa , showing the manifest $\IZ_2$ duality covariance
of the RG flows.

This treatment also allowed us to examine at higher orders
the relation between
RG/duality consistency, which is motivated from a 2d field theory
point of view, and duality invariance of the string background
effective action, which is motivated from a string theory
point of view. Insofar as our consistency relations turn out to
be stronger requirements than duality invariance of the background
effective action, we have proposed that such a consistency
requirement be elevated to a basic principle, to be enforced at
each order in perturbation theory. From it, duality
invariance of the background effective action follows once the
simpler requirement is made that the measure factor $\sqrt{g}e^{-2\f}$
also be invariant under duality.

Finally, we investigated the consistency between RG flows and T-duality
explicitly at two-loop order, for a restricted class of backgrounds.
The fact that the beta functions are modified in going one order
higher suggests that the form T-duality transformations take should
also be modified at each higher order. Borrowing from the work of
Tseytlin \tseytlina , we considered one of two sets of modified
duality transformations for the particular backgrounds
in question, and verified that
the consistency conditions are exactly satisfied
also at this order.
The picture that emerges at two-loop order is that, although
formulas become more complicated due to the perturbative
corrections they receive, the consistency between RG flows
and T-duality survives these complications unscathed. RG flows
continue to flow covariantly with duality, and fixed points
of a model are mapped to fixed points of its dual. Although
at this order the duality symmetry is still $\IZ_2$, we have
not been able to express the consistency relations in a form
which expresses this symmetry manifestly.

The ultimate goal in our endeavor is to understand in more precise
terms the nature of quantum corrections to T-duality transformations
for generic backgrounds, and fully understand the ``hierarchy''
(if indeed there is one) between the requirement of duality
invariance of the background effective action, and the requirement
of duality covariance of the RG flows in the sigma model.
Even at two-loop order this is an ambitious task. In order to
progress in that direction, it seems to us the next step would
naturally be to consider classes of backgrounds which are more
encompassing than the one considered here, if perhaps not
entirely generic at first. For instance, the introduction of
antisymmetric background fields is particularly interesting,
as it would bring in for the first time the scheme dependence
present in the \oaa\ beta functions. It would furthermore allow
for a comparison with the special cases provided by WZW models,
where some exact results are known (cf. \kiritsis\ and related
work). The particular class of backgrounds we initially have in
mind contain a generic metric and no torsion in the original target,
corresponding to a block diagonal metric plus torsion in the dual.
We have already initiated such an investigation.

We finally note that work similar in spirit to what we have presented
here has also been done in entirely different contexts, namely
for lattice spin systems, and the quantum Hall effect \burgess .
In the string/sigma model context,
for $D\!+\!1$-dimensional backgrounds with $D$ isometries, the 
preservation of duality symmetry, in this case $O(D,D)$, at two-loop
order has recently been considered in \meiss .
T-duality has also been studied for massive sigma models, and thus
away from conformal points, in \hewson\ and, in the case of open 
strings, its interplay with the RG has been considered in \dorn.

\appendix{A}{Kaluza-Klein Reduction}

For the sake of the assiduous reader who would like to reproduce
our results, we list below all
quantities relevant for our computations. We write a generic background
metric $g_{\m\n}$ as in \metric , and the components of the
antisymmetric
background tensor $b_{\m\n}$ as $b_{0i}\equiv w_i$ and $b_{ij}$. In this
notation, barred quantities refer to the metric $\bar{g}_{ij}$.

\item{1)} {\it Inverse metric}:  $g^{00}\!=\!1/a+v_iv^i,\
g^{0i}\!=\!-v^i,\ g^{ij}\!=\!\bar{g}^{ij}$.  On decomposed tensors,
indices $i,j,\ldots$ are raised and lowered with the metric
$\bar{g}_{ij}$ and its inverse. With the metric decomposition
\metric\ we also have $\det g= a\det\bar{g}$.\medskip

\item{2)} {\it Connection coefficients}:
\eqn\connection{\eqalign{
\G^0_{00}&={a\over2}v^ia_i\ ,\ \G^0_{i0}={a\over2}\left[ {a_i\over a}+
v^ja_jv_i+v^jF_{ji}\right]\ ,\cr
\G^i_{00}&=-{a\over2}a^i\ ,\ \G^i_{0j}=-{a\over2}\left[ F^i_{\
j}+a^iv_j\right]\ ,\cr
\G^0_{ij}&=-\bar{\G}^k_{ij}v_k+\half (\pa_iv_j+\pa_jv_i+a_iv_j+a_jv_i)-
{a\over2}v^k\left[ v_jF_{ik}+v_iF_{jk}-a_kv_iv_j\right]\ ,\cr
\G^i_{jk}&=\bar{\G}^i_{jk}+{a\over2}\left[ v_jF_k^{\ i}+v_kF_j^{\ i}
-a^iv_jv_k\right]\ ,}}
where $a_i\!=\!\pa_i\ln a\ ,\ F_{ij}\!=\!\pa_iv_j-\pa_jv_i$.
\medskip

\item{3)} {\it Ricci tensor}:
\eqn\ricci{\eqalign{
R_{00}&=-{a\over2}\left[ \bar{\nabla}_ia^i+\half a_ia^i-{a\over2}F_{ij}
F^{ij}\right]\ ,\cr
R_{0i}&=v_iR_{00}+{3a\over4}a^jF_{ij}+{a\over2}\bar{\nabla}^jF_{ij}\ ,\cr
R_{ij}&=\bar{R}_{ij}+v_iR_{0j}+v_jR_{0i}-v_iv_jR_{00}-\half
\bar{\nabla}_ia_j
-{1\over4}a_ia_j-{a\over2}F_{ik}F_j^{\ k}\ .}}\medskip

\item{4)}{\it Riemann tensor}:
\eqn\riemann{\eqalign{
R_{i0k0}&= -{a\over2}\left({1\over2}a_{i}a_{k}+\bar{\nabla}_{i}a_{k}+
{a\over2}F_{i}^{\ l}F_{lk}\right)\ ,\cr
R_{ijk0}&= v_{j}R_{i0k0}-v_{i}R_{j0k0}-{a\over2}\bar{\nabla}_{k}F_{ij}
-{a\over2}\left( a_{k}F_{ij}+{1\over2}a_{j}F_{ik}-{1\over2}a_{i}F_{jk}
\right)\ ,\cr
R_{ijkm}&= \bar{R}_{ijkm}+R_{ijk0}v_{m}+R_{jim0}v_{k}+R_{mkj0}v_{i}
+R_{kmi0}v_{j}\ ,\cr
&\ -R_{m0j0}v_{i}v_{k}+R_{k0j0}v_{i}v_{m}-R_{k0i0}v_{j}v_{m}
+R_{m0i0}v_{j}v_{k}\ ,\cr
&\ -{a\over4}\left(F_{im}F_{kj}+F_{ki}F_{mj}+2F_{ji}F_{mk}
\right). }}\medskip

\item{5)} {\it Torsion}:
\eqn\torsion{\eqalign{ H_{0ij}&=-\pa_iw_j+\pa_jw_i\equiv -G_{ij}\ ,\cr
H_{ijk}&=\pa_ib_{jk}+\pa_jb_{ki}+\pa_kb_{ij}\ ,}} and all other
components vanish. For the one-loop beta function the following
quantities are needed:\medskip
\eqn\torsionsq{\eqalign{ H_{0\m\n}H_0^{\ \m\n}&=G_{ij}G^{ij}\ ,\cr
H_{0\m\n}H_i^{\ \m\n}&=-2G_{ij}G^{jk}v_k-H_{ijk}G^{jk}\ ,\cr
H_{i\m\n}H_j^{\ \m\n}&=2\left({1\over a}+v_mv^m\right) G_i^{\ k}G_{jk}
-2v^kv^mG_{ik}G_{jm}+2H_{km(i}G_{j)}^{\ k}v^m\cr &+H_{ikm}H_j^{\ km}\ ,}} and
\eqn\divtorsion{\eqalign{ \nabla_\m H^\m_{\
0i}&=\bar{\nabla}^jG_{ji}-aG_{ij}F^{jk}v_k+\half G_{ij}a^j
-{a\over2}F^{jk}\left( H_{ijk}+v_iG_{jk}\right)\ ,\cr \nabla_\m H^\m_{\
ij}&=\bar{\nabla}^k\left( H_{kij}+v_kG_{ij}\right) -\half \left[ G_i^{\
k}\bar{\nabla}_{(k}\ v_{j)} -G_j^{\ k}\bar{\nabla}_{(k}\ v_{i)}
\right]-{a\over2}v_{[i}H_{j]km}F^{km}\cr &+v_{[i}G_{j]k}\left(
a^k-aF^{km}v_m\right)+\half a^kH_{kij}+\half v_ma^mG_{ij}-\half
F_{[i}^{\ k}G_{j]k}\ ,}} where $[ij]=ij-ji$ and $(ij)=ij+ji$.\medskip
                              
\item{6)} {\it Dilaton terms}:
\eqn\dilaton{\eqalign{
\nabla_0\pa_0\f&={a\over2}a^i\pa_i\f\ ,\cr
\nabla_0\pa_i\f&={a\over2}\left(F^j_{\ i}+a^jv_i\right) \pa_j\f\ ,\cr
\nabla_i\pa_j\f&=\bar{\nabla}_i\pa_j\f -{a\over2}\left( v_iF_j^{\ k}+
v_jF_i^{\ k}-a^kv_iv_j\right)\pa_k\f\ .}}

\item{(7)} {\it Tangent space geometrical tensors}:\medskip

When referred to the tangent space, the Ricci tensor becomes
\eqn\riccitangent{\eqalign{
R_{\hat{0}\hat{0}}&=-{1\over2}
\left[ \bar{\nabla}_ia^i+\half a_ia^i-{a\over2}F_{ij}
F^{ij}\right]\ ,\cr
R_{\hat{0}\a}&=\bar{e}^{\ i}_{\a}
\left[ {3\sqrt{a}\over4}a^jF_{ij}+{\sqrt{a}\over2}\bar{\nabla}^jF_{ij}
\right]\ ,\cr
R_{\a\b}&=\bar{e}^{\ i}_{\a}\bar{e}^{\ j}_{\b}
\left[\bar{R}_{ij}-\half \bar{\nabla}_ia_j
-{1\over4}a_ia_j-{a\over2}F_{ik}F_j^{\ k}\right]\ ,}}
where $\bar{e}^{\ i}_{\a}$ is the inverse vielbein for the metric
$\bar{g}_{ij}$. Likewise, the Riemann tensor is
\eqn\riemanntangent{\eqalign{
R_{\a \hat{0} \b \hat{0}}&= -{1\over2}\bar{e}^{\ i}_{\a}\bar{e}^{\
j}_{\b}
\left({1\over2}a_{i}a_{j}+\bar{\nabla}_{i}a_{j}
+{a\over2}F_{i}^{\ s}F_{sj}\right)\ ,\cr
R_{\a \b \g \hat{0}}&= -{\sqrt{a}\over2}\bar{e}^{\ i}_{\a}\bar{e}^{\
j}_{\b}
\bar{e}^{\ k}_{\g}\left( a_{k}F_{ij}+{1\over2}a_{j}F_{ik}
-{1\over2}a_{i}F_{jk}+\bar{\nabla}_{k}F_{ij}\right)
\ ,\cr
R_{\a \b \g \d}&=
\bar{e}^{\ i}_{\a}\bar{e}^{\ j}_{\b}\bar{e}^{\ k}_{\g}
\bar{e}^{\ m}_{\d}\left( \bar{R}_{ijkm}-{a\over4}
( F_{im}F_{kj}+F_{ki}F_{mj}+2F_{ji}F_{mk})\right) . }}

\listrefs

\end